\documentclass[runningheads]{llncs}
\usepackage{graphicx}
\usepackage{multirow}
\begin{document}
\title{Demystifying Pythia: A Survey of ChainLink Oracles Usage on Ethereum}
%
%
\author{Mudabbir Kaleem \and
Weidong Shi}
\authorrunning{M. Kaleem and W. Shi}
%
\institute{University of Houston, Texas, USA\\
\email{\{mkaleem,wshi3\}@uh.edu}}
\maketitle              
\begin{center}
Published in the proceedings of the 25th International Conference on Financial Cryptography and Data Security's 1st Workshop on Decentralized Finance (FC DeFi 2021).
\end{center}

\begin{abstract}
Smart contracts are dependent on oracle systems for their adoption and usability. We perform an empirical study of oracle systems' usage trends and adoption metrics to provide better insight into the health of the smart contract ecosystem. We collect ChainLink usage data on the Ethereum network using a modified Ethereum client and running a full node. We analyze the collected data and present our findings and insights surrounding the usage trends, adoption metrics, oracle pricing and service quality associated with ChainLink on the Ethereum network. We infer that ChainLink's usage and growth are dominated by the DeFi ecosystem and for its demand for decentralized price feeds.

\end{abstract}
\section{Introduction}
Since the launch of the Ethereum~\cite{buterin2014ethereum} network in 2015, smart contracts~\cite{szabo1997formalizing} have become one of the central features of blockchain-based systems. Although initially limited in usage to token control and on-chain data access, smart contracts today are rapidly expanding their domain of applications~\cite{kehrli2016blockchain} due to the availability of oracles~\cite{al2020trustworthy}. Oracles provide the interface between the blockchain's isolated execution environment and external off-chain data sources, enabling smart contracts to retrieve and post real-world data and events. Consequently, the potential utility and future mass adoption of smart contract platforms is inextricably tied to the oracle service providers within the ecosystem.

Bearing that in mind, the motivation of this study was to survey oracle usage in the smart contract ecosystem. Currently, different projects like ChainLink~\cite{chainlink}, Provable~\cite{provable} and Augur~\cite{augur} are offering third party oracle services to smart contracts. These projects have adapted a decentralized approach for collecting and aggregating oracle data, thereby addressing "the oracle problem"~\cite{egberts2017oracle} of having centralized points of failures in blockchain environments. For our survey, we target ChainLink, which evidently captures the majority share of the oracle middleware market at the time of writing. To establish this, we surveyed the top forty DeFi projects by market capitalization~\cite{coinmardefi} and found all among them which had a use case for external oracles to be using ChainLink except two projects. ChainLink provides a comprehensive list of their project integrations on their website~\cite{chainlinkecosys} and it includes major DeFi projects such as Aave, Ampleforth, Chiliz, Polygon, Kyber Network and 0x among others. Although ChainLink provides its oracle services over multiple chains, we concern our study with ChainLink oracle usage on Ethereum since it is the most widely adopted smart contract platform at this time. We believe that ChainLink oracle usage on Ethereum represents the significant bulk of oracle traffic on smart contract platforms. Our study finds that Chainlink's growth and usage is strongly centered around the DeFi ecosystem where a few projects have been responsible for most of the oracle service traffic for price feeds. We also show that Chainlink's price feeds feature has seen a steady growth since its inception whereas the external API feature has seen negligible traffic. The oracle traffic statistics and trends provided by this survey can be used to gauge the adoption and health of the smart contract ecosystem in general.  At the time of writing, we are not aware of any other formal study providing oracle usage insights in the smart contract environment.

\section{ChainLink Overview}
ChainLink is an oracle service provider for smart contracts that is currently live on three platforms: Ethereum, Binance Chain and the Matic Network. ChainLink went live in May 2019 and is currently the most popular oracle service provider for smart contracts. ChainLink maintains a decentralized oracle network and aggregates data from multiple oracle nodes on the network to provide data feeds that do not rely on a single oracle node or data source\cite{chainlinkDoc}. ChainLink employs an ERC-20 and ERC-677 compliant token called LINK which is used by oracle consumers to pay the oracle nodes for data provision. ChainLink currently provides three features for consumer smart contracts on the Ethereum mainnet.

\subsubsection{Price Feeds:} are a ChainLink feature to provide different market prices and conversion rates data in the blockchain environment for usage by smart contracts. ChainLink achieves this by having a decentralized price feed for each of these data points, which is fed price data through multiple oracle nodes using different sources. This is implemented by having an aggregator contract for each feed on-chain which is fed data by multiple oracle nodes through their interface contracts. The feed aggregator contract then aggregates all the nodes' answers to provide a final answer to any consumer contract via public Solidity functions. Consumers of the price feeds data call these aggregator contracts when the data is desired. The ChainLink documentation lists the aggregator contract addresses for the available price feeds\cite{chainlinkPriceFeeds}. The price feeds are sponsored by various projects and currently available for public usage without any LINK token charge.
\subsubsection{External APIs:} is a ChainLink feature that allows smart contracts in the blockchain environment to perform external API calls through ChainLink oracle nodes. These API calls can be HTTP Get Requests on the web or other APIs provided by the oracle node for different use cases. ChainLink API requests are currently handled 1:1 by an oracle and ChainLink currently does not provide decentralization benefits by default for API calls although a user might implement it on their own. The consumers of ChainLink's API feature have to pay their request servicing oracle node in LINK tokens for the service. The cost varies depending on the node and the nature of the request but is around 0.1 LINK on average and the highest being 1 LINK at the time of writing. Commonly used public API endpoints are available as "jobs" in ChainLink which allows user to only specify the job ID and not having to specify the URL, format etc. This makes the consumer side code more succinct and the implementation easier.
\subsubsection{Verifiable Random Numbers (VRF):} is a ChainLink feature to provide verifiable random number generation functionality on-chain. ChainLink achieves this by having off-chain random number verifier contracts which verify the randomness of the number generated by an oracle node in response to a consumer request. VRF feature allows for provable random numbers, which protects the consumer from attacks even if the node servicing the request has been compromised.

\section{Study Design}
\subsection{Data Collection}
For both the Price Feeds and the External APIs we collected data from the launch of ChainLink mainnet in May 2019 up till the end of October 2020 (Ethereum block 11167816). The VRF feature data was not collected and is not part of this study since it only went live at the end of October 2020 and the resulting data was insufficient for a formal study.
\subsubsection{Modified Ethereum Client:} For collecting the Price Feed usage data, we looked at the price feed addresses available on the ChainLink website~\cite{chainlinkPriceFeeds}. There were 88 price feed addresses at the time of writing which are proxy aggregator addresses. ChainLink has also, since its launch, made upgrades to the aggregator contracts. The current version of aggregators are labeled as v3. We used the wayback machine web archives~\cite{machine2015internet} to retrieve old aggregator addresses and had a total of 169 addresses for our study (88 v3, 80 v2, 1 v1). The ChainLink team also later provided us with historical addresses which we used to verify our list. For capturing the price feed data we could not use the Web3 API since all price feed consumer requests were direct calls or "internal transactions". Hence we modified the Golang Ethereum client code to log data when internal function calls were made to these 169 addresses. We captured the block number, calling address, opcode, value and input data parameters for these internal calls to these addresses and stored them in a local MySQL database.

\subsubsection{Ethereum Full Node and Web3:}
For collecting data related to ChainLink API usage we used the Web3 API with an Ethereum full node that we ran locally. ChainLink implements the API feature using the CallAndTransfer() functionality of the ERC-677 token standard. Every time a consumer requests an oracle, it generates a ChainlinkRequest event and sends the LINK to the oracle node along with data describing the API to fetch, the job ID, the format of the output, the callback address and function which the oracle must respond to and other data if required. The oracle node interface contract generates an OracleRequest event upon receiving the LINK and data and the external node listens to this event. It responds with the result after some time and makes a transaction to the callback function with the data response. The consumer contract then raises a ChainlinkFulfilled event. We use the Web3 APIs to capture these events and extract the required data which includes: the block number of the request, the requesting address, the oracle node requested, the job ID specified, the callback function and address provided, the LINK token paid, the ChainLink request ID, the request transactions hash, any additional data provided, the response block number, the response and the response transaction hash. We store the results in our local MySQL database for all such oracle service request-response cycles on ChainLink.

We used Etherscan~\cite{team2017etherscan} to verify various samples of our collected data to ensure that our data collection process was performed correctly. 

\subsection{Study Objectives}
The study was aimed at providing insights into the usage of ChainLink oracles on Ethereum. For this purpose we looked at the following five aspects:
\begin{itemize}
\item \textbf{Oracle usage trends and demographics}
\item \textbf{Oracle Adoption}
\item \textbf{Oracle Pricing}
\item \textbf{Oracle Servicing Delays}
\end{itemize}

\section{Results}
\subsection{Usage Trends and Demographics}
After the data collection was completed and the required data was populated into our MySQL server, we had the quantitative information summarized in \tablename ~\ref{tab:overall}. A total of 2,717,049 API requests were made to Oracles during the entire duration of our study and in total 2,409,074 price feed calls were made to ChainLink's public price feed contracts for fetching the market place data. Although the numbers appear encouraging at first sight, upon further investigation, we found that 99.75\% of API requests to ChainLink oracle nodes were made by ChainLink price feed aggregator addresses themselves. This is because prior to the v3 aggregator release in August 2020~\cite{chainlinkdevcomm}, all price feed aggregator contracts made API requests to oracle nodes to fetch prices. After removing these API requests, we are only left with 6634 API requests performed on ChainLink for the entire 18 month period! We also see that the number of distinct users that made use of these features is very low.

\begin{figure}
\begin{center}
\includegraphics[width=3.0in,angle=0]{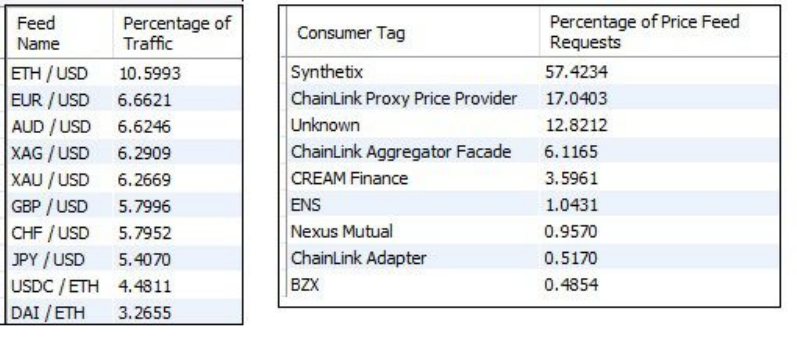}
\caption{Leaderboards: Price Feeds attracting the most traffic and projects generating the most price feed traffic.}
\label{fig:firstfig}
\end{center}
\end{figure}

Next, we present a list of the most popular price feeds based on their share of the historical price-feed traffic in \figurename~\ref{fig:firstfig}. We also present the corresponding consumer projects/contracts of these price feeds ordered by their share of the historical price-feed traffic. To get the corresponding projects/contracts, we grouped the most regular consumer addresses (Top 26 addresses, which represent more than 90\% of all price-feed traffic) by their public tags available on Etherscan~\cite{team2017etherscan}. Our results show that Synthetix~\cite{synthetix}, which is a blockchain-based derivatives trading platform, is responsible for more than 47\% of the historic price feed traffic. If we subtract ChainLink's internal traffic from the numbers, Sythetix's share of the historical price feed traffic rises to 75\%.

\begin{table}
\scriptsize
\caption{Price Feeds and API: collected data summary.}
\label{tab:overall}
\centering
\begin{tabular}{|l|l|l|l|l|}
\hline
Feature                & Total Requests & \begin{tabular}[c]{@{}l@{}}Excluding ChainLink\\Internal Requests\end{tabular} & \begin{tabular}[c]{@{}l@{}}Distinct Caller\\/Consumer Addresses\end{tabular} & \begin{tabular}[c]{@{}l@{}}Distinct Callee\\Addresses (Price-\\Feeds/Oracle Nodes)\end{tabular}  \\ 
\hline
\textbf{Price Feeds}   & 2409074        & N/A                                                                            & 294                                                                                  & 129                                                                                              \\ 
\hline
\textbf{External APIs} & 2717049        & 6634                                                                           & 271                                                                                  & 159                                                                                              \\
\hline
\end{tabular}
\end{table}

\vspace{-5pt}

\begin{figure}
\begin{center}
\includegraphics[width=5.0in,angle=0]{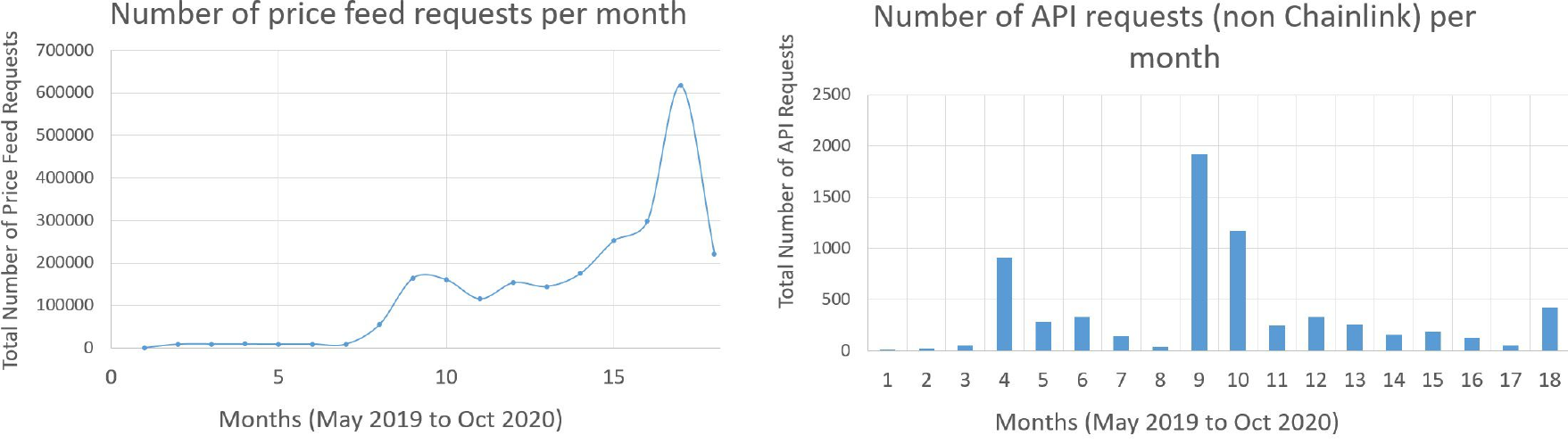}
\caption{Number of price feed and API requests on ChainLink by month.}
\label{fig:Reqspermonth}
\end{center}
\end{figure}
\vspace{-5pt}

\subsection{Oracle Adaption in the Market}
\begin{figure}
\begin{center}
\includegraphics[width=5.0in,angle=0]{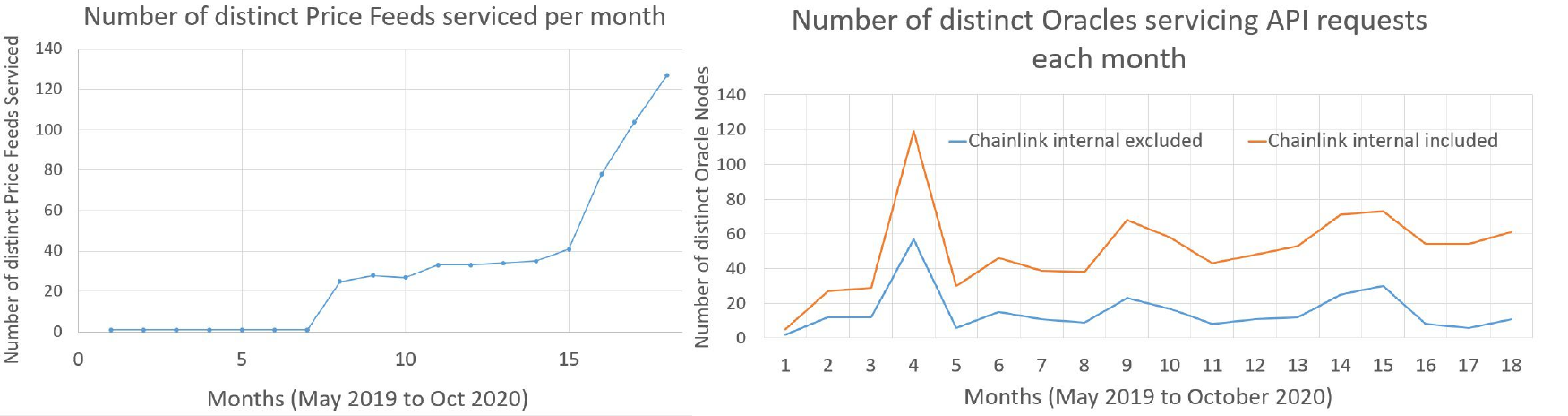}
\caption{Number of distinct Price Feeds serviced and active Oracles by month.}
\label{fig:dispricefeedpermonth}
\end{center}
\end{figure}

To study ChainLink oracles' adaption trends in the market, we look at the historical data for the average number of price-feed and API requests made to ChainLink oracles per month \figurename~\ref{fig:Reqspermonth}. Plotting the data, we can see that the price-feed feature appears to be far more popular among users and has been rapidly gaining more traffic volume. The API feature does not appear to have a large demand among the users. We believe that this can be attributed to the fact that most projects and use-cases are able to fulfill their data needs using the ChainLink provided price feeds and do not have to employ a custom API.
We also show in \figurename~\ref{fig:dispricefeedpermonth} that ChainLink has continuously increased the number of price feeds being offered to users. The increase in price feed offerings has kept up with the increase in adaption as evidenced in these figures. In contrast to the price-feeds, Oracle nodes have not seen a marked increase in the variety of API calls and jobs being requested.

\vspace{-5pt}

\subsection{Oracle Pricing}
\begin{figure}
\begin{center}
\includegraphics[width=5.0in,angle=0]{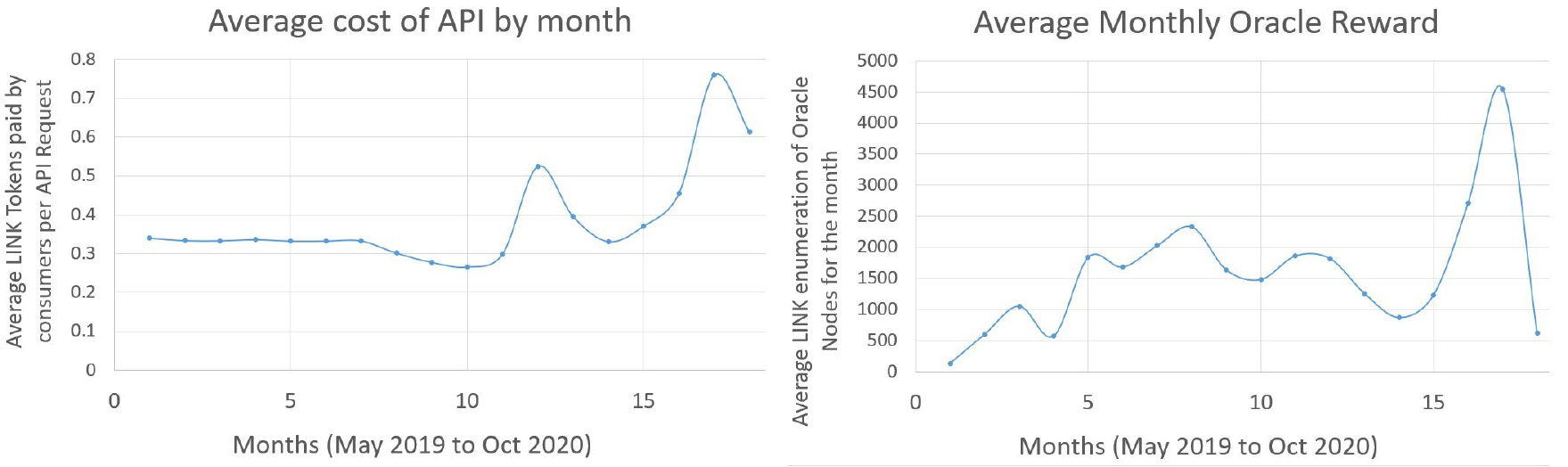}
\caption{Average cost of a single API request and the Average fees collected in LINK by oracle nodes.}
\label{fig:CostsOracles}
\end{center}
\end{figure}

ChainLink is currently providing the price feeds feature to all smart contract users on the Ethereum chain without cost. These price feeds are sponsored by various blockchain projects using these feeds in their contracts. A user does need to pay an Oracle node in LINK token if they make a direct API request. The current cost of ChainLink API usage varies and can be as high as 1 LINK depending on the oracle and the data being requested. We look at the historical price paid for running a single API request in \figurename~\ref{fig:CostsOracles}. We also look at the historical average income which the data providing oracle nodes from these requests. We see the average LINK paid for oracle requests on ChainLink is increasing of late, and that coupled with the increase in the LINK token price is bound to discourage the use of oracle APIs for trivial use cases.

\subsection{Oracle Servicing Delays}
\begin{figure}
\begin{center}
\includegraphics[width=5.0in,angle=0]{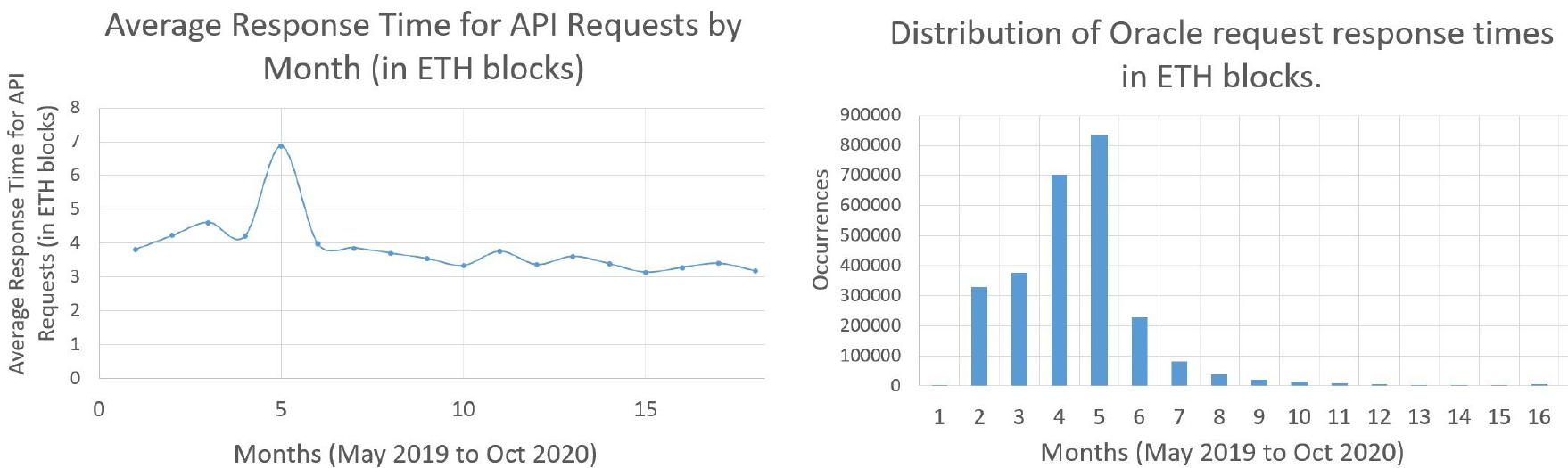}
\caption{Average response time and response time distribution for API requests.}
\label{fig:responseTimesFig}
\end{center}
\end{figure}

Different smart contract use cases require their oracle service requests to be processed within a time constraint. For the wide adoption of smart contracts, it is essential that the oracle system is able to service time-critical requests. We analyze our available API data in \figurename~\ref{fig:responseTimesFig} to determine the historical average delay experience on ChainLink API requests. Due to a small number of outliers, the average obtained was around six hundred blocks. After filtering out these outliers and only keeping the requests that were serviced within one hundred blocks, we obtained the data shown in our figures. We can see that for ChainLink oracles most API requests are serviced within the next four to five blocks with the historical average block delay being close to four Ethereum blocks which corresponds to roughly one minute.

\vspace{-5pt}

\vspace{-5pt}

\section{Analysis and Conclusion}
Based on our analysis of the collected data, we obtained the following important insights regarding Chainlink usage on Ethereum in particular and the trends for Oracle systems in general:
\begin{itemize}
\item	The number of individual users of the ChainLink platform is not very high. Currently, it is mostly being used by DeFi(Decentralized Finance) projects and applications to provide market prices to its contracts. This is perhaps indicative of a trend in the smart contract ecosystem in general.
\item	Currently, a single DeFi project, Synthetix has been responsible for almost 75\% of the historic price-feed traffic in the ChainLink network (given that we ignore ChainLink's self-generated traffic). Synthetix uses various commodity and currency ratio feeds on ChainLink which are among the feeds that have serviced the most traffic. This dominance of Synthetix related traffic might fade with ChainLink increasingly integrating with new projects. 
\item	The data shows that there is currently not a big market of people wanting to use oracles to connect smart contracts to the external world for trivial use cases. Whether it is the genuine lack of market demand for these applications or whether high Ethereum gas prices and ChainLink API fees discourage people from doing so will require further investigation.   
\item	While ChainLink's API feature has not seen increased use with the rise of DeFi, ChainLink's price feeds have seen increasing usage since the project's launch. ChainLink has also managed to provide an increasing variety of price feeds to cater to the demands of new DeFi projects integrating with ChainLink.
\item	The rising average API cost seen on the network might be attributed to the increased LINK token price which forces people to only make Oracle API usage for non-trivial cases.
\item	The average response time of ChainLink's API feature is seen to remain steady between 4 and 5 blocks which might not be good enough for time-sensitive applications.
\end{itemize}

In conclusion, at the time of this study, the ChainLink ecosystem on the Ethereum network appears to be driven purely by DeFi's demand for decentralized market price feeds~\cite{liu2020look}. In the coming future, it would be interesting to see if Oracle platforms like ChainLink take initiatives to attract other segments of users or tailor themselves more towards fulfilling the needs of the growing DeFi market.
\paragraph*{Acknowledgements} 
The authors warmly thank the ChainLink team for sharing historical price feed addresses with us for cross-verification.

%
%
%

\bibliographystyle{splncs04}
\bibliography{references}
\end{document}